\documentclass{elsarticle}
\usepackage{algorithm}
\usepackage{algorithmic}
\pdfoutput=1
\usepackage{graphicx}

\begin{document}

\begin{frontmatter}

\title{Accelerating numerical solution of Stochastic Differential Equations with CUDA}

\author[sil]{M. Januszewski, M. Kostur}

\address[sil]{Institute of Physics, University of Silesia, 40-007 Katowice, Poland}

\begin{abstract}
Numerical integration of stochastic differential equations is
commonly used in many branches of science. In this paper we present
how to accelerate this kind of numerical calculations with popular NVIDIA Graphics
Processing Units using the CUDA programming environment. We address general aspects of
numerical programming on stream processors and illustrate them by two
examples: the noisy phase dynamics in a Josephson junction and the noisy
Kuramoto model.  In presented cases the measured speedup can be as high
as $675\times$ compared to a typical CPU, which corresponds to several
billion integration steps per second.  This means that calculations
which took weeks can now be completed in less than one hour. This
brings stochastic simulation to a completely new level, opening
for research a whole new range of problems which can now be solved
interactively.
\end{abstract}

\end{frontmatter}

\section{Introduction}

The numerical integration of stochastic differential equations
(SDEs) is a valuable tool for analysis of a vast diversity of problems
in physics, ranging from equilibrium transport in molecular motors
\cite{Reimann200257}, phase dynamics in Josephson junctions
\cite{kostur:104509,kautz1999}, stochastic resonance
\cite{RevModPhys.70.223} to dissipative particle dynamics
\cite{DPD1997} to finance \cite{McLeish}. Stochastic simulation, as it
is referred to as, is specially interesting when the dimensionality of
the problem is larger than three, and in that case it is often the
only effective numerical method. A prominent example of this is the
stochastic variation of molecular dynamics: Brownian dynamics.

Direct stochastic simulations require a significant computational
effort, and therefore merely a decade ago have been used mostly as
validation tools. The precise numerical results in theory of
low-dimensional stochastic problems were coming from solutions of
the corresponding Fokker-Planck equations. Many different sophisticated,
but often complicated, tools have been applied: spectral methods
\cite{PhysRevLett.76.1166,PhysRevE.59.1417,PhysRevE.61.6320}, finite
element methods \cite{kostur2002} and numerical path integrals
\cite{Yu_PI2004,naess:021126}.

Stochastic simulation gained acceptance due to its
straightforward implementation and generic robustness with respect to
different sorts of problems.  The continuous increase of the
efficiency of available computer hardware has been acting in favour of
stochastic simulation, making it increasingly more popular. The recent
evolution of computer architectures towards multiprocessor and
multicore platforms also resulted in improved performance of
stochastic simulation. Let us note that in the case of a low-dimensional
system, stochastic simulation often uses ensemble
averaging to obtain the values of observables, which in turn is an
example of a so-called ,,embarrassingly parallel problem'' and it can,
though with embarrassment, directly benefit from a parallel
architecture. In other cases, mostly where a large number of
interacting subsystems are investigated, the implementation of the
problem on a parallel architecture is less trivial, but still possible.

The recent emergence of techniques collectively known as
general-purpose computing on graphics processing units (GPUs) has caused
a breakthrough in computational science.  The current state of the art
GPUs are now capable of performing computations
at a rate of about 1 TFLOPS per single silicon chip. It must be stressed
that 1 TFLOPS is a performance level which only in 1996 was
achievable exclusively by huge and expensive supercomputers such as the ASCI
Red Supercomputer (which had a peak performance of $1.8$ TFLOPS \cite{asciired})
The numerical simulations of SDEs can easily benefit from the parallel GPU
architecture. This however requires careful redesign of the employed
algorithms and in general cannot be done automatically. In this paper
we present a practical introduction to solving SDEs on NVIDIA GPUs
using Compute Unified Device Architecture (CUDA) \cite{cuda0} based on two
examples: the model of phase diffusion in a Josephson junction and
the Kuramoto model of coupled phase oscillators.

The paper is organized as follows: first, we briefly introduce the
features and capabilities of the NVIDIA CUDA environment and describe
the two physical models, then we present the implementation of
stochastic algorithms and compare their efficiency with a
corresponding pure-CPU implementation executed on an Intel Core2 Duo E6750 processor.
We also provide the source code \cite{progs} of three small example programs:
{\em PROG1}, {\em PROG2}, and {\em PROG3}, which demonstrate the techniques described in the paper.
They can easily be extended to a broad range of problems
involving stochastic differential equations.

\section{The CUDA environment}

\begin{figure}[tpb]
	\centering
	\includegraphics{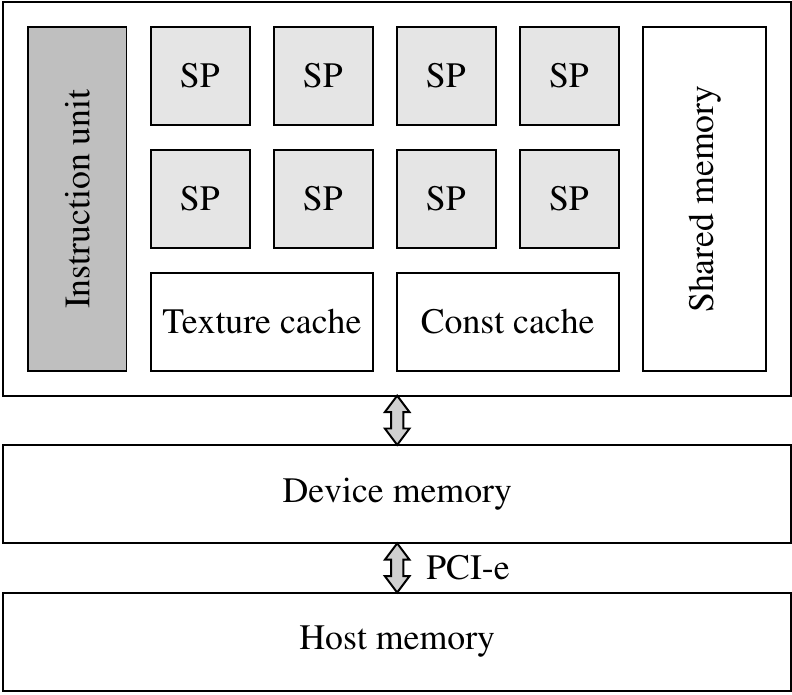}
	\caption{A schematic view of a CUDA streaming multiprocessor with 8 scalar processor cores.}
	\label{fig:cuda}
\end{figure}

CUDA (Compute Unified Device Architecture) is the name of a general purpose parallel
computing architecture of modern NVIDIA GPUs.  The name \emph{CUDA}
is commonly used in a wider context to refer to not only the hardware architecture of the GPU, but also
to the software components used to program that hardware.  In this sense, the CUDA
environment also includes the NVIDIA CUDA compiler and the system drivers and libraries
for the graphics adapter.

From the hardware standpoint, CUDA is implemented by organizing the GPU
around the concept of a streaming multiprocessor (SM).  A modern NVIDIA
GPU contains tens of multiprocessors.  A multiprocessor consists of 8 scalar
processors (SPs), each capable of executing an independent thread (see Fig.~\ref{fig:cuda}).  The multiprocessors
have four types of on-chip memory:
\begin{itemize}
	\item a set of 32-bit registers (local, one set per scalar processor)
	\item a limited amount of shared memory (16 kB for devices having Compute Capability 1.3 or lower,
		shared between all SPs in a MP)
	\item a constant cache (shared between SPs, read-only)
	\item a texture cache (shared between SPs, read-only)
\end{itemize}
The amount of on-chip memory is very limited in comparison to the total global memory
available on a graphics device (a few kilobytes vs hundreds of megabytes).  Its advantage
lies in the access time, which is two orders of magnitude lower than the global memory
access time.

The CUDA programming model is based upon the concept of a \emph{kernel}.
A kernel is a function that is executed multiple times in parallel, each
instance running in a separate thread.  The threads are organized into
one-, two- or three-dimensional blocks, which in turn are organized into
one- or two-dimensional grids.  The blocks are completely independent of each other and
can be executed in any order. Threads within a block however are guaranteed
to be run on a single multiprocessor.  This makes it possible for them to
synchronize and share information efficiently using the on-chip memory
of the SM.

In a device having Compute Capability 1.2 or higher, each multiprocessor
is capable of concurrently executing 1024 active threads~\cite{cudaPG2.1}.
In practice, the number of concurrent threads per SM is also limited by the amount
of shared memory and it thus often does not reach the maximum allowed value.

The CUDA environment also includes a software stack.  For CUDA v2.1, it
consists of a hardware driver, system libraries implementing the CUDA API,
a CUDA C compiler and two higher level mathematical libraries (CUBLAS and CUFFT).
CUDA C is a simple extension of the C programming language, which includes
several new keywords and expressions that make it possible to distinguish
between host (i.e. CPU) and GPU functions and data.

\section{Specific models}

In this work, we study the numerical solution of stochastic
differential equations modeling the dynamics of Brownian
particles.  The two models we concentrate upon are of particular
interest in many disciplines and illustrate the flexibility of
the employed methods of solution.

The first model describes a single Brownian particle moving in a
symmetric periodic potential $V(x) = \sin (2 \pi x)$ under the
influence of a constant bias force $f$ and a periodic unbiased driving
with amplitude $a$ and frequency $\omega$:
\begin{equation}
\label{eq:josephson}
\ddot{x} + \gamma \dot{x} = -V'(x) + a \cos(\omega t) + f + \sqrt{2 \gamma k_B T} \xi(t)
\end{equation}
where $\gamma$ is the friction coefficient and $\xi(t)$ is a zero-mean
Gaussian white noise with the auto-correlation function $\langle \xi(t)
\xi(s) \rangle = \delta(t - s)$ and noise intensity $k_B T$.

Equation~\ref{eq:josephson} is known as the Stewart-McCumber model
\cite{kautz1999} describing phase differences across a Josephson
junction. It can also model a rotating dipole in an external field, a
superionic conductor or a charge density wave.  It is particularly
interesting since it exhibits a wide range of behaviors, including
chaotic, periodic and quasi-periodic motion, as well as the recently
detected phenomenon of absolute negative mobility
\cite{machura:040601,Reimann2007}.

The second model we analyze is that of $N$ globally interacting overdamped
Brownian particles, with the dynamics of the $i$-th particle
described by:
\begin{eqnarray}
\label{eq:kuramoto}
\gamma \dot{x_i} = \omega_i + \sum_{j=1}^{N} K_{ij} \sin(x_j - x_i)
+ \nonumber \\
\sqrt{2 \gamma k_B T} \xi_i(t), i = 1, \ldots, N
\end{eqnarray}

This model is known as the Kuramoto model \cite{acebron:137} and is used as a simple
paradigm for synchronization phenomena.  It has found applications in
many areas of science, including neural networks, Josephson junction
and laser arrays, charge density waves and chemical oscillators.

\section{Numerical solution of SDEs}

Most stochastic differential equations of practical interest cannot be
solved analytically, and thus direct numerical methods have to be used
to obtain the solutions. Similarly as in the case of ordinary
differential equations, there is an abundance of methods and
algorithms for solving stochastic differential equations. Their
detailed description can be found in references:
\cite{PhysRevA.40.3381,Mannella2002,PhysRevA.26.1589,PhysRevA.38.5938,PhysRevA.45.600,Kloeden}.

Here, we present the implementation of a standard stochastic
algorithm on the CUDA architecture in three distinctive cases:
\begin{enumerate}
\item Multiple realizations of a system are simulated, and an
  ensemble average is performed to calculate quantities of interest.
  The large degree of parallelism inherent in the problem makes it possible to fully exploit
  the computational power of CUDA devices with tens of multiprocessors
  capable of executing hundreds of threads simultaneously. The example
  system models the stochastic phase dynamics in a Josephson junction
  and is implemented in program {\em  PROG1} (the source code is available in \cite{progs}).
\item The system consists of $N$ globally interacting particles. In
  each time step $N^2$ interaction terms are calculated. The example
  algorithm is named {\em PROG2} and solves the Kuramoto
  model~(Eq.~\ref{eq:kuramoto}.)
\item The system consists of $N$ globally interacting particles as in the
  previous case but the interaction can be expressed in terms of
  a parallel reduction operation, which is much more efficient than {\em PROG2}.
  The example algorithm in {\em PROG3} also solves the Kuramoto
  model~(Eq.~\ref{eq:kuramoto}.)
\end{enumerate}

We will now outline the general patterns used in the solutions of
all models.  We start with the model of a single Brownian particle,
which will form a basis upon which the solution of the more general
model of $N$ globally interacting particles will be based.

\subsection{Ensemble of non-interacting stochastic systems}

\begin{algorithm*}
\caption{A CUDA kernel to advance a Brownian particle by $m \cdot \Delta t$ in time.}
\label{alg:kernsingle}
\begin{algorithmic}[1]
\STATE local $i \leftarrow blockIdx.x \cdot blockDim.x + threadIdx.x$
\STATE load $x_i$, $v_i$ and system parameters $\{par_{ji}\}$ from global memory and store them in local variables
\STATE load the RNG seed $seed_i$ and store it in a local variable
\FOR{$s = 1$ to $m$}
\STATE generate two uniform variates $n_1$ and $n_2$
\STATE transform $n_1$ and $n_2$ into two Gaussian variates
\STATE advance $x_i$ and $v_i$ by $\Delta t$ using the SRK2 algorithm
\STATE local $t \leftarrow t_0 + s \cdot \Delta t$
\ENDFOR
\STATE save $x_i$, $v_i$ and $seed_i$ back to global memory
\end{algorithmic}
\end{algorithm*}

\begin{algorithm*}
	\caption{The Stochastic Runge-Kutta algorithm of the 2nd order (SRK2) to integrate
	$\dot{x} = f(x) + \xi(t)$, $\langle \xi(t) \rangle = 0$,
	$\langle \xi(t) \xi(s) \rangle = 2 D \delta (t - s)$.}
	\label{alg:srk2}
\begin{algorithmic}[1]
\STATE $F_1 \leftarrow f(x_0)$
\STATE $F_2 \leftarrow f(x_0 + \Delta t F_1 + \sqrt{2D \Delta t} \psi$) \COMMENT{with $\langle \psi \rangle = 0$, $\langle \psi^2 \rangle = 1$}
\STATE $x(\Delta t) \leftarrow x_0 + \frac{1}{2} \Delta t (F_1 + F_2) \sqrt{2D \Delta t} \psi$
\end{algorithmic}
\end{algorithm*}

For the Josephson junction model described by Eq.~\ref{eq:josephson}
we use a single CUDA kernel, which is responsible for advancing the
system by a predefined number of timesteps of size $\Delta t$.

We employ fine-grained parallelism -- each path is calculated in a
separate thread.  For CUDA devices, it makes sense
to keep the number of threads as large as possible.  This enables the
CUDA scheduler to better utilize the available computational power
by executing threads when other ones are waiting for global memory transfers
to be completed~\cite{cudaPG2.1}.
It also ensures that the code will execute efficiently on new GPUs,
which, by the Moore's law, are expected to be capable of
simultaneously executing exponentially larger numbers of threads.  We
have found that calculating $10^5$ independent realizations is enough
to obtain a satisfactory level of convergence and that further
increases of the number of paths do not yield better results (see
Fig.~\ref{fig:anmsup}).

In order to increase the number of threads, we structured our code so
that Eq.~\ref{eq:josephson} is solved for multiple values of
the system parameters in a single run.  The default setup calculates
trajectories for $100$ values of the amplitude parameter $a$.  This
makes it possible to use our code to efficiently analyze the behavior
of the system for whole regions of the parameter space $\{a, \omega, \gamma\}$.

Multiple timesteps are calculated in a single kernel invocation to
increase the efficiency of the code.  We observe that usually only
samples taken every $M$ steps are interesting to the researcher
running the simulation, the sampling frequency $M$ being chosen so
that the relevant information about the analyzed system is retained.
In all following examples $M = 100$ is used.  It should be noted that
the results of the intermediate steps do not need to be copied to the
host (CPU) memory.  This makes it possible to limit the number of
global memory accesses in the CUDA threads.  When the kernel is
launched, path parameters $x$, $v = \dot{x}$ and $a$ are loaded from
the global memory and are cached in local variables.  All calculations
are then performed using these variables and at the end of the kernel
execution, their values are written back to the global memory.

Each path is associated with its own state of the random number generator (RNG),
which guarantees independence of the noise terms between different
threads.  The initial RNG seeds for each thread are chosen randomly using a 
standard integer random generator available on the host system.
Since CUDA does not provide any random number generation routines by default,
we implemented a simple xor-shift RNG as a CUDA device function.  In our
kernel, two uniform variates are generated per time step and
then transformed into Gaussian variates using the Box-Muller transform.
The integration is performed using a Stochastic Runge-Kutta scheme of the
2nd order, which uses both Gaussian variates for a single time step.

\begin{figure}[tpb]
	\centering
	\includegraphics[width=12cm]{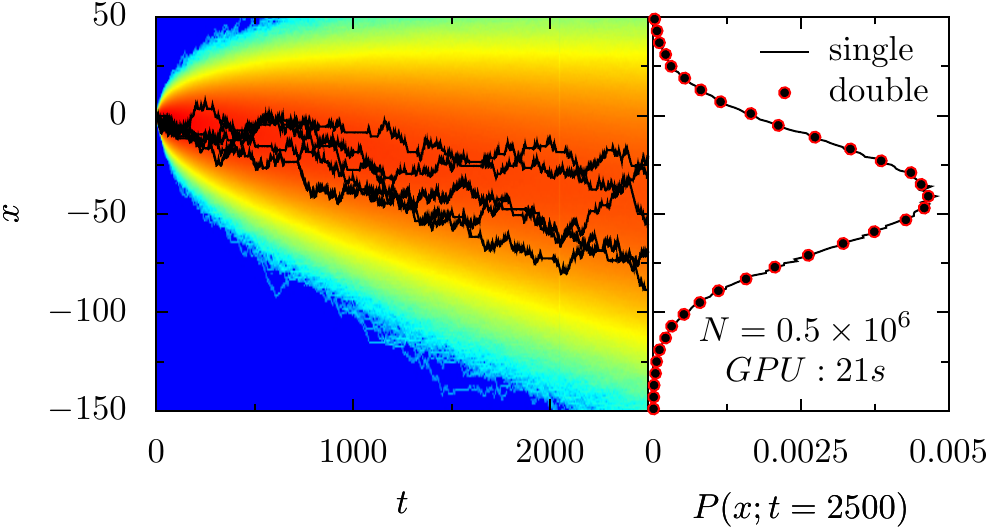}
	\caption{The ensemble of $524288$ Brownian particles,
          modeling the noisy dynamics of phase in a Josephson
          junction described by Eq. \ref{eq:josephson} is simulated
          for time $t\in(0,2000 \frac{2 \pi}{\omega})$ with time step
          $\Delta t = 0.01 \frac{2 \pi}{\omega}$. On the left panel
          sample trajectories are drawn with black lines and the
          background colors represent the coarse-grained
          (averaged over a potential period) density of particles in
          the whole ensemble. The right panel shows the coarse-grained
          probability distribution of finding a particle at time
          $t=2000\frac{2 \pi}{\omega}$ obtained by means of a
          histogram with $200$ bins.  The histogram is calculated
          with both single and double precision on a GPU with Compute Capability v1.3.
		  The same calculation has also been performed on the CPU
		  but their identical results are not presented for clarity
		  purposes.  The total simulation times were: \textbf{20 seconds}
		  and 13 minutes on NVIDIA Tesla 1060C when using single and double precision
		  floating-point arithmetics, respectively. The CPU-based version of
          the same algorithm needed over three hours. Used parameters: $a
          = 4.2$, $\gamma = 0.9$, $\omega = 4.9$, $D_0 = 0.001$,
          $f=0.1$ correspond to the anomalous response regime
          (cf. \cite{machura:040601}).}
	\label{fig:precision}
\end{figure}

In the example in Fig.~\ref{fig:precision} we present the results
coming from the simultaneous solution of $N=2^{19}=524288$ independent
Eqs.~\ref{eq:josephson} for the same set of parameters. The total
simulation time was less than $20$~seconds. In this case the CUDA platform
turns out to be extremely effective, outperforming the CPU by a factor of
$675$. In order to highlight the amount of computation, let us note that
the size of the intermediate file with all particle positions used
for generation of the background plot was about $30$~GB.

\subsection{$N$ globally interacting stochastic systems}

\begin{algorithm*}
	\caption{The \textbf{AdvanceSystem} CUDA kernel.}
	\label{alg:advancesystem}
\begin{algorithmic}[1]
\STATE local $i \leftarrow blockIdx.x \cdot blockDim.x + threadIdx.x$
\STATE local $mv \leftarrow 0$
\STATE local $mx \leftarrow x_i$
\FORALL{tiles}
\STATE local $tix \leftarrow threadIdx.x$
\STATE $j \leftarrow tile \cdot blockDim.x + threadIdx.x$
\STATE shared $sx_{tix} \leftarrow x_j$
\STATE synchronize with other threads in the block
\FOR{$k = 1$ to $blockDim.x$}
\STATE $mv \leftarrow mv + \sin(mx - sx_k)$
\ENDFOR
\STATE synchronize with other threads in the block
\ENDFOR

\STATE $v_i \leftarrow mv$
\end{algorithmic}
\end{algorithm*}

\begin{figure}[tpb]
\centering
	\includegraphics[width=7cm]{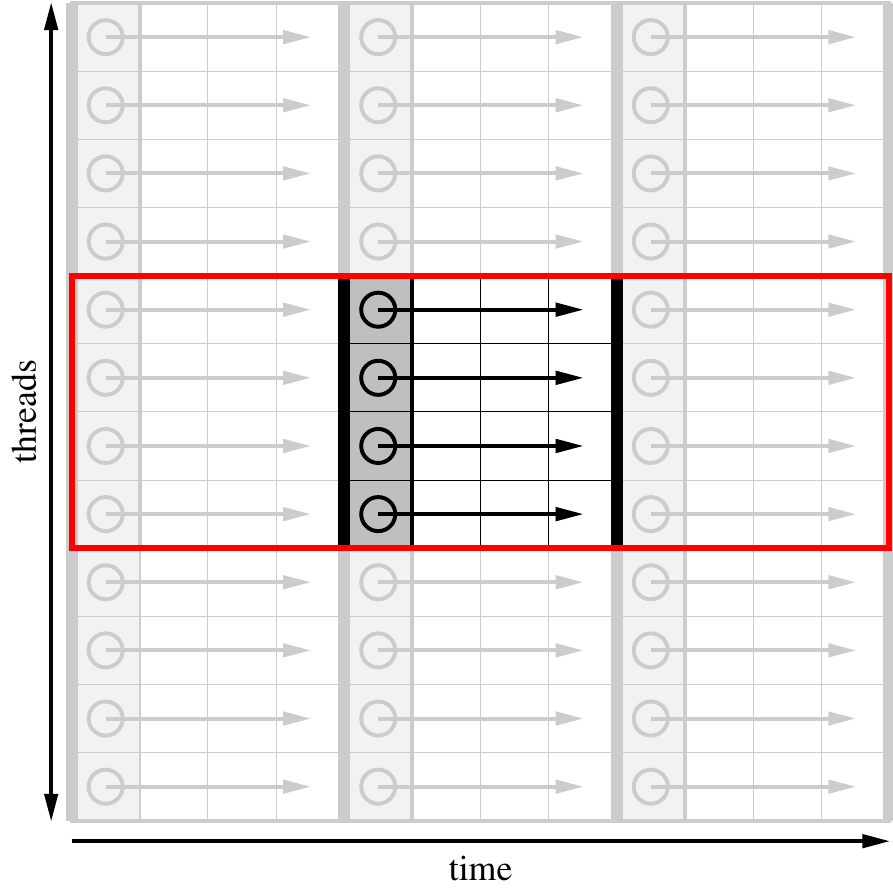}
	\caption{All-pairs interaction of 12 particles calculated using the tile-based approach with 9 tiles of size 4x4.
		The chosen number of particles and the size of the tiles are made artificially low for illustration purposes only.
		A small square represents the computation of a single particle-particle interaction term.  The highlighted
		part of the schematic depicts a single tile.  The bold lines represent synchronization points where data
		is loaded into the shared memory of the block.  The filled squares with circles represent the start of
		computation for a new tile.  Threads in the red box are executed within a single block.}
	\label{fig:tile}
\end{figure}

For the general Kuramoto model described by Eqs.~\ref{eq:kuramoto} or
other stochastic systems of $N$ interacting particles,
the calculation of $\mathcal{O}(N^2)$ interaction terms for
all pairs $(x_j, x_i)$ is necessary in each integration step.
In this case the program {\em PROG2} is
split into two parts, implemented as two CUDA kernels launched
sequentially.  The first kernel, called \textbf{UpdateRHS} calculates
the right hand side of Eq.~\ref{eq:kuramoto} for every $i$.  The second
kernel \textbf{AdvanceSystem} actually advances the system by a single
step $\Delta t$ and updates the positions of all particles.  In our
implementation the second kernel uses a simple first-order Euler
scheme.  It is straightforward to modify the program to implement
higher-order schemes by interleaving calls to the \textbf{UpdateRHS}
kernel with calls to kernels implementing the sub-steps of the scheme.

The \textbf{UpdateRHS} kernel is organized around the concept of
\emph{tiles}, introduced in \cite{GPUgems3:nbody}.
A tile is a group of $T$ particles interacting with another group of $T$ particles.
Threads are executed in blocks of size $T$ and each block is always processing
a single tile.  There is a total of $N/T$ blocks in the grid.  The $i$-th thread
computes the interaction of the $i$-th particle with all other particles.

The execution proceeds as follows.  The $i$-th thread loads the position of the
$i$-th particle and caches it as a local variable.  It then loads the position
of another particle from the current tile, stores it in shared memory and synchronizes
with other threads in the block.  When this part is completed, the positions of
all particles from the current tile are cached in the shared memory.  The computation
of the interaction is then commenced, with the $i$-th thread computing the interaction
of the $i$-th particle with all particles from the current tile.  Afterwards, the
kernel advances to the following tile, the positions stored in shared memory
are replaced with new ones, and the whole process repeats.

This approach might seem wasteful since it computes exactly $N^2$ interaction
terms, while only $(N-1) N / 2$ are really necessary for a symmetric
interaction.  It is however very efficient, as it minimizes global memory
transfers at the cost of an increased number of interaction term computations.
This turns out to be a good trade-off in the CUDA environment, as global
memory accesses are by far the most costly operations, taking several
hundred clock cycles to complete.  Numerical computations are comparatively
cheap, usually amounting to just a few clock cycles.

\begin{figure}[tpb]
	\centering
	\includegraphics[width=8cm]{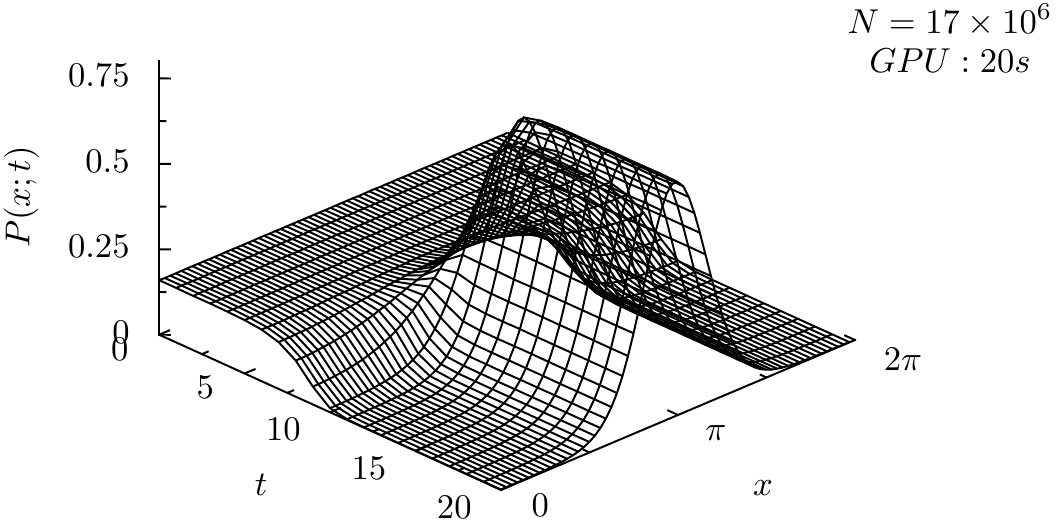}
	\caption{An example result of the integration of the Kuramoto system
    (Eq. ~\ref{eq:kuramoto}). The time evolution of the probability density
    $P(x;t)$ is shown for $\omega_i=0$, $K_{ij}=4$, $T=1$. The density is
    a position histogram of $2^{24}$ particles. The total time of
    simulation was approximately $20$ seconds using the single precision
    capabilities of NVIDIA Tesla C1060.}
	\label{fig:kuramoto3d}
\end{figure}

The special form of the interaction term in the Kuramoto model when
$K_{ij}=K=\mathrm{const}$, allows us to significantly simplify the
calculations.  Using the identity:
\begin{eqnarray}
	\label{eq:SC}
	\sum_{j=1}^{N} \sin(x_j - x_i) = \nonumber \\
		\cos(x_i) \sum_{j=1}^{N} \sin(x_j) - \sin(x_i) \sum_{j=1}^{N} \cos(x_j) &&
\end{eqnarray}
we can compute two sums: $\sum_{j=1}^{N} \sin(x_j)$ and
$\sum_{j=1}^{N} \cos(x_j)$ only once per integration step, which
has a computational cost of $\mathcal{O}(N)$. The calculation of the sum of a vector
of elements is an example of the vector reduction operation, which can be performed
very efficiently on the CUDA architecture. Various methods of
implementation of such an operation are presented in the sample code included
in the CUDA SDK 2.1~\cite{cudaSDK}. The integration of the Kuramoto system
taking advantage of Eq.~\ref{eq:SC} and using a simple form of a
parallel reduction is implemented in {\em PROG3}.

In Fig.~\ref{fig:kuramoto3d} we present a solution of the classical
Kuramoto system described by Eqs.~\ref{eq:kuramoto} for parameters as
in Fig.~10 of the review paper \cite{acebron:137}. In this case we apply
the program {\em PROG3} which makes use of the relation from Eq.~\ref{eq:SC}.  The number of
particles $N=2^{24} \approx 16.8 \cdot 10^6$ and the short simulation time
clearly demonstrate the power of the GPU for this kind of problems.

\section{Note on single precision arithmetics}

The fact that the current generation of CUDA devices only implements
single precision operations in an efficient way is often considered
a significant limitation for numerical calculations.  We have found
out that for the considered models this does not pose a problem.
Figure~\ref{fig:precision} presents sample paths and position
distribution functions of a Brownian particle whose dynamics is
determined by Eq.~\ref{eq:josephson} (colored background on
the left panel and right panel). Let us note that we present
coarse-grained distribution functions where the position is averaged
over a potential period by taking a histogram with bin size being exactly equal
to the potential period.  We observe that the use of single precision
floating-point numbers does not significantly impact the obtained
results.  Results obtained by single precision calculations even after
a relatively long time $t=2000\frac{2\pi}{\omega}$ differ from their
double precision counterparts only up to the statistical error, which in
this case can be estimated by the fluctuations of the relative particle number in
a single histogram bin.  Since in the right panel of
Fig.~\ref{fig:precision} we have approximately $10^4$ particles in one
bin, the error is of the order of $1\%$.  If time-averaged quantities
such as the asymptotic velocity $\langle \langle v \rangle \rangle = \lim_{t \to \infty} \langle v(t) \rangle$
are calculated, the differences are even less pronounced.  However, the single
and double precision programs produce different individual
trajectories as a direct consequence of the chaotic nature of the
system given by Eq.~\ref{eq:josephson}.  Moreover, we have noticed that
even when changing between GPU and CPU versions of the same program, the
individual trajectories diverged after some time.  The difference
between paths calculated on the CPU and the GPU, using the same
precision level, can be explained by differences in the floating-point
implementation, both in the hardware and in the compilers.

When doing single precision calculations special care must be taken to ensure
that numerical errors are not needlessly introduced into the calculations.
If one is used to having all variables defined as double precision floating-point
numbers, as is very often the case on a CPU, it is easy to forget that operations
which work just fine on double precision numbers might fail when single precision
numbers are used instead.  For instance, consider the case of keeping track of time in
a simulation by naively increasing the value of a variable $t$ by a constant $\Delta t$
after every step.  By doing so, one is bound to hit a problem when $t$ becomes large
enough, in which case $t$ will not change its value after the addition of a small value $\Delta t$,
and the simulation will be stuck at a single point in time.  With double precision numbers
this issue becomes evident when there is a difference of 17 orders of magnitude
between $t$ and $\Delta t$.  With single precision numbers, a 8-orders-of-magnitude
difference is enough to trigger the problem.  It means that if, for instance, $t$ is $10^5$ and
$\Delta t$ is $10^{-4}$, the addition will no longer work as expected. $10^5$ and $10^{-4}$ are
values not uncommon in simulations of the type we describe here, hence the need
for extra care and reformulation of some of the calculations so that very large and very
small quantities are not used at the same time.  In our implementations, we avoided the
problem of spurious addition invariants by keeping track of simulation time modulo
the system period $2 \pi / \omega$.  This way, the difference between $t$ and $\Delta t$
was never large enough to cause any issues.

\section{Performance evaluation}

In order to evaluate the performance of our numerical solution of Eqs.~\ref{eq:josephson}
and~\ref{eq:kuramoto}, we first implemented Algs.~\ref{alg:advancesystem}
and~\ref{alg:kernsingle} using the CUDA Toolkit v2.1.  We then translated the CUDA code
into C++ code by replacing all kernel invocations with loops and removing unnecessary
elements (such as references to shared memory, which does not exist on a CPU).

We used the NVIDIA CUDA Compiler (NVCC) and GCC 4.3.2 to compile the CUDA code and
the Intel C++ Compiler (ICC) v11.0 for Linux to compile the C++ version.
We have determined through numerical experiments that enabling floating-point
optimizations significantly improves the performance of our programs (by a factor
of $7$ on CUDA) and does not affect the results in a quantitative or qualitative way.  We
have therefore used the \texttt{-fast -fp-model fast=2} ICC options and
\texttt{--use\_fast\_math} in the case of NVCC.

\begin{figure}[tpb]
	\centering
	\includegraphics[width=12cm]{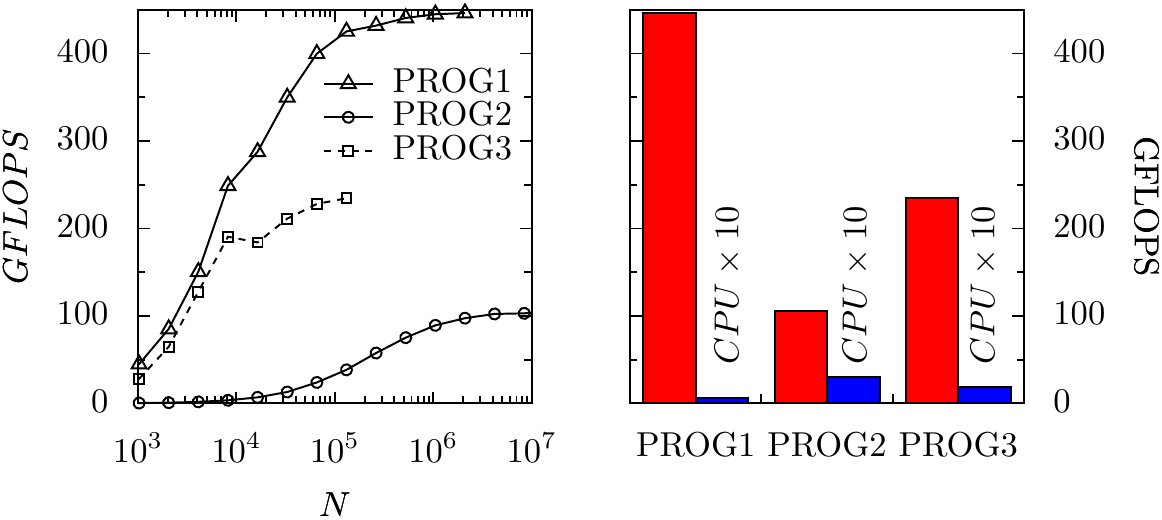}
	\caption{(Left panel) Performance estimate for the programs \emph{PROG1}-\emph{PROG3} as a function of the number of particles $N$.
	(Right panel) Performance estimate for the programs \emph{PROG1}-\emph{PROG3} on an Intel Core2 Duo E6750 CPU and
	NVIDIA Tesla C1060 GPU. We have counted $79$, $44+6 N$ and  $66$ operations per one integration step of programs
	\emph{PROG1}, \emph{PROG2} and \emph{PROG3}, respectively.}
	\label{fig:anmsup}
\end{figure}


All tests were conducted on recent GNU/Linux systems using the following hardware:
\begin{itemize}
	\item for the CPU version: Intel Core2 Duo E6750  @ 2.66GHz and 2 GB RAM (only a single core was used for the calculations)
	\item for the GPU version: NVIDIA Tesla C1060 installed in a system with Intel Core2 Duo CPU E2160  @ 1.80GHz and 2 GB RAM
\end{itemize}

Our tests indicate that
speedups of the order of 600 and 100 are possible for the
models described by Eqs.~\ref{eq:josephson}
and~\ref{eq:kuramoto}, respectively.  The performance gain is
dependent on the number of paths used in the
simulation. Figure~\ref{fig:anmsup} shows that it increases
monotonically with the number of paths, and then saturates at a number
dependent on the used model: $450$ and $106$ GFLOPS for the
Eqs.~\ref{eq:josephson} and~\ref{eq:kuramoto}, respectively (which
corresponds to speedups: $675$ and $106$).  The saturation point
indicates that for the corresponding number of particles the full
computational resources of the GPU are being exploited.

The problem of lower performance gain for small numbers of particles could
be rectified by dividing the computational work between threads in a different
way, i.e.~by decreasing the amount of calculations done in a single thread, while
increasing the total number of threads.  This is a relatively straightforward
thing to do, but it increases the complexity of the code.  We decided not
to do it since for models like \ref{eq:josephson} and \ref{eq:kuramoto} one
is usually interested in calculating observables for whole ranges of system
parameters.  Instead of modifying the code to run faster for lower number of
paths, one can keep the number of paths low but run the simulation for multiple
system parameters simultaneously, which results in a higher number of threads.

\section{Conclusions}

In this paper we have demonstrated the suitability of a parallel
CUDA-based hardware platform for solving stochastic differential
equations. The observed speedups, compared to CPU versions, reached an
astonishing value $670$ for non-interacting particles and $120$ for
a globally coupled system.  We have also shown that for this kind of
calculations single precision arithmetics poses no
problems with respect to accuracy of the results, provided that
some kind of operations, such as adding small and large numbers, are
avoided.

The availability of cheap computer hardware which is over two orders
of magnitude faster clearly announces a new chapter in high
performance computing. Let us note that the development of stream processing
technology for general-purpose computing has just started and its potential is
surely not yet fully revealed. In order to take advantage of the new
hardware architecture, the software and its algorithms must be
substantially redesigned.

\section{Appendix: Estimation of FLOPS}

We counted the floating-point operations performed by the kernels in
our code, and the results in the form of the collective numbers of
elementary operations are presented in Table \ref{table:flops}. The
number of MAD (Multiply and Add) operations can vary, depending on how
the compiler processes the source code.  For the purposes of our
performance estimation, we assumed the most optimistic version.  A more
conservative approach would result in a lower number of MADs, and
correspondingly a higher total number of GFLOPS.

\begin{table}
\centering
\caption{Number of elementary floating-point operations performed per
one time step in the \textbf{AdvanceSystem} kernel for Eq.~\ref{eq:josephson}.}
\label{table:flops}
\begin{tabular}{ | c |c| c |c| }
\hline
  count  &  type & FLOPs & total\newline FLOPs \\ \hline \hline
 22             & multiply, add        & 1                   & 22 \\ \hline
 11             & MAD                  & 1                   & 11 \\ \hline
 2              & division             & 4                   & 8 \\ \hline
 3              & sqrt                 & 4                   & 12 \\ \hline
 1              & $\sin$               & 4                   & 4 \\ \hline
 5              & $\cos$               & 4                   & 20 \\ \hline 
 1              & $\log$               & 2                   & 2 \\ \hline 
 \multicolumn{3}{|r|}{TOTAL:}                        & \textbf{79} \\ \hline
\end{tabular}
\end{table}

The amount of FLOPs for functions such as sin, log, etc. is based on
\cite{cudaPG2.1}, assuming $1$ FLOP for elementary
arithmetical operations like addition and multiplication and scaling
the FLOP estimate for complex functions proportionately to the number
of processor cycles cited in the manual. The numbers of floating-point
operations are summarized in Table \ref{table:flops}.

On a Tesla C1060 device our code {\em PROG1} evaluates
$6.178 \cdot 10^9$ time steps per second.  The cost of each
time step is $79$ FLOPs, which implies that the overall
performance estimate accounts for $490$ GFLOPS.

In the case of {\em PROG2} the number of operations per one integration
step depends on the number of particles $N$. A similar operation
count as the one presented in Table \ref{table:flops} resulted in the formula $44+6 N$
FLOPs per integration step.

\bibliography{stochastic}
\bibliographystyle{cpc}

\end{document}